\documentclass{PoS}

\def\siml{{\ \lower-1.2pt\vbox{\hbox{\rlap{$<$}\lower6pt\vbox{\hbox{$\sim$}}}}\ }}
\def\simg{{\ \lower-1.2pt\vbox{\hbox{\rlap{$>$}\lower6pt\vbox{\hbox{$\sim$}}}}\ }}

\def \als {\alpha_{\mathrm{s}}}

\def \m2   {\mu^{2 \epsilon}}
\def\lQ{\Lambda_{\rm QCD}}
\def\als{\alpha_{\rm s}} 
\def\bea{\begin{eqnarray}}
\def\eea{\end{eqnarray}}
\def\be{\begin{equation}}
\def\ee{\end{equation}}

\title{Current Topics in Heavy Quarkonium Physics}

\ShortTitle{BORMIO2011}

\author{\speaker{Nora Brambilla}\\
 Physik-Department, Technische Universit\"at M\"unchen,
James-Franck-Str. 1, 85748 Garching, Germany\\
        E-mail: \email{nora.brambilla@ph.tum.de}}

\abstract{I review  some recent progress, open puzzles and  future opportunities in heavy quarkonium physics  in the framework 
of effective field theories.}

\FullConference{XLIX International Winter Meeting on Nuclear Physics, BORMIO2011\\
		January 24-28, 2011\\
		Bormio, Italy}

\begin{document}

\section{Quarkonium}

Heavy Quarkonia   are systems composed by two heavy quarks, with mass $m$ larger than the 
``QCD  confinement scale'' $\lQ$, so that $\als(m) \ll 1$ holds.
They are  multiscale systems.
 Being nonrelativistic, quarkonia are characterized by another small parameter, 
the heavy-quark velocity $v$, 
 ($v^2 \sim 0.1$ for $b\bar{b}$, $v^2 \sim 0.3$ for $c\bar{c}$,  
$v^2 \sim 0.01$ for $t\bar{t}$),
and by a hierarchy of energy scales: $m$ (hard),
the  relative momentum $p \sim m v$ (soft),
 and the binding energy $E \sim m v^2$ (ultrasoft).
For energy scales close to $\lQ$, perturbation theory breaks down  
and one has to rely on nonperturbative 
methods. Regardless of this, the nonrelativistic hierarchy $m \gg m v \gg m v^2$ 
persists also below the $\lQ$ threshold. While the hard scale  $m$ is always  larger than 
$\lQ$, different  situations may arise for the other two scales 
depending on the considered quarkonium system.
The  soft scale, proportional to the  inverse quarkonium radius $r$,
may be a perturbative ($\gg \lQ$) or a nonperturbative scale ($\sim \lQ$) depending 
on the physical system in consideration. 
The first case is likely to happen only for the lowest charmonium and 
bottomonium states. We do not have direct
information on the radius of the quarkonia systems, and thus the
attribution of  some of the lowest bottomonia and charmonia states 
to the perturbative or the nonperturbative soft regime is at the
moment still ambiguous \cite{GarciaiTormo:2005bs}.
Only for $t\bar{t}$ threshold states  the ultrasoft
scale  may  be considered still perturbative.

All these quarkonium scales get entangled in a typical amplitude involving a quarkonium
observable. In particular, quarkonium annihilation 
and production take place  at the scale $m$,  quarkonium binding takes place at the scale
$mv$, which is the typical momentum exchanged inside the bound state, while 
very low-energy gluons and light quarks (also called ultrasoft degrees of freedom)  
live long enough that a bound state has time to form and, therefore, are sensitive to the 
scale $mv^2$. Ultrasoft gluons are responsible for phenomena similar to the  the Lamb shift in QCD.

The existence of many scales  in quarkonium makes it a unique system to 
study complex environments. Quarkonium probes all the regimes of QCD, from the 
high energy region, where an expansion in the the coupling constant is 
possible, to the low energy region, where nonperturbative effects dominate. It 
probes also the intermediate region between the two regimes.
In particular for quarkonium systems with a very small radius the 
interaction turns out to be purely perturbative 
while for systems with a large radius 
with respect to the confinement scale the interaction turns out to be 
nonperturbative.
Therefore quarkonium  is an ideal and to some extent unique laboratory 
where our understanding of nonperturbative QCD and its interplay with 
perturbative QCD may be tested in a controlled framework.
The fact that the quarkonium interaction is dominated by the glue
contribution makes it a particularly precious  system to test models of 
physics beyond the Standard Model (BSM)  where a treatment of confinement  and nontrivial low energy
configurations is introduced. The large mass, the clean and known decays mode make quarkonium an ideal
probe of new physics in some well defined  window of parameters of beyond the 
Standard Model (BSM), in particular for some dark matter candidates search 
\cite{Brambilla:2010cs,Brambilla:2004wf,SanchisLozano:2010zza, McElrath:2005bp}.  In the complex environment of heavy ion collisions
quarkonium suppression constitutes  a unique probe  of deconfinement and 
quark gluon plasma formation  \cite{Brambilla:2010cs,Brambilla:2004wf}. 
The different radius of the different 
quarkonia states induces the phenomenon of sequential suppression, 
allowing to use quarkonium as  a kind of thermometer for the measurement of the temperature of the formed 
medium. On similar ground quarkonium may constitute  a  special probe
to be used in the study  of a  nuclear medium \cite{Lutz:2009ff}.

The diversity, quantity and accuracy of the data collected at experiments in the last few 
years is impressive and includes
data on quarkonium formation  from BES  and BESIII at BEPC and BEPC2, 
KEDR  at VEPP-4M, and CLEO-III and CLEO-c at CESR;
clean samples of charmonia produced in B-decays, in photon-photon
fusion and in initial state radiation from the B-meson factory
experiments BaBar at SLAC and Belle at KEK, including the unexpected
observation of large associated $(c\overline{c})(c\overline{c})$ production;
heavy quarkonia production from gluon-gluon fusion in $p\bar{p}$
annihilations at 2~TeV from the CDF and D0 experiments at Fermilab;
charmonia production in heavy-ion collisions from the PHENIX and STAR
experiments at RHIC.
These experiments  have largely  operated as quarkonium factories producing large data sample on quarkonium
spectra,  decays  and production  with high statistics.
New states and exotics, 
new production mechanisms, new transitions and  unexpected states of an exotic nature 
have been observed. The study of quarkonium in media has also undergone crucial developments,
the suppression of quarkonium production in the hot medium remaining one of  the 
cleanest and most relevant probe of deconfined matter.
New data are already copiously coming from LHC experiments and new facilities will become
operational (Panda at GSI, a much higher luminosity B
factory at KEK, possibly a SuperB) adding challenges and opportunities to this research field.

From the theory point of view,  effective field theories  (EFTs)  as HQET (Heavy Quark Effective Theory), NRQCD (Non Relativistic 
QCD),  pNRQCD (potential Non Relativistic QCD),  SCET (Soft Collinear Effective Theory)...,  for the description of
 quarkonium processes have been newly developed and are being  developed, providing a unifying description as well as a solid and versatile 
tool giving well-defined, model independent and precise predictions
\cite{Brambilla:2010cs, Brambilla:2004wf, Brambilla:2004jw}.
They rely on one  hand on  high order perturbative calculations and on 
the other hand on lattice simulations, the recent  progress in both fields 
having added a lot to the theory reach.

The progress in our understanding of nonrelativistic EFTs makes it possible to move beyond 
phenomenological models (at least for states below threshold) and to provide in this case 
a systematic description  inside QCD of heavy-quarkonium physics. On the other hand, the recent 
progress in the measurement of several heavy-quarkonium observables makes it meaningful 
to address the problem of their precise theoretical determination.
In this situation quarkonium  becomes a special system to advance our theoretical 
understanding of the strong interactions, also in special environments (e.g. quarkonium in media)
and in several production mechanisms, as well as  our control of some parameters of the Standard
Model

The International Quarkonium Working Group  (QWG) (www.qwg.to.infn.it)
created in 2002 has supplied an adequate platform for discussion and common research work between theorists 
and experimentalists, producing also two large reviews of  state of the art, open problems,  perspective and opportunities 
of quarkonium physics  in 2010 and 2004 \cite{Brambilla:2010cs}. In particular at the end of \cite{Brambilla:2010cs} 
is presented a list of 65 priorities in experiments and in theory. Some of the results appeared in  the last few months
already challenged such list. In the following I will address some aspects of this  research field  which is at the moment in great evolution.

\section{Theory: the effective field theory description}

A hierarchy of EFTs may be constructed by systematically integrating out 
modes associated to  high energy scales not relevant for quarkonium \cite{Brambilla:2004jw}.
Such integration  is made  in a matching procedure enforcing 
the  equivalence between QCD and the EFT at a given 
order of the expansion in $v$.
The EFT  realizes a factorization at the Lagrangian level between 
the high energy contributions, encoded into the  matching coefficients, and 
 the low energy contributions, carried by the dynamical degrees of freedom.
Poincar\'e symmetry remains  intact in a nonlinear realization at the level of the nonrelativistic
 (NR) EFT and imposes exact relations among the matching coefficients  \cite{Brambilla:2003nt}.

By integrating out the hard modes Nonrelativistic QCD (NRQCD) is obtained
 \cite{Caswell:1985ui,Bodwin:1994jh,Manohar:1997qy},
 making explicit at the Lagrangian level the expansions in $m v/m$ and $mv^2/m$.
 The effective Lagrangian is organized as an expansion in $1/m$  and $\als(m)$.
 It is is similar to HQET, but with a different power counting and
accounts also for contact interactions 
between quarks and antiquark pairs (e.g. in decay processes), hence having a wider 
set of operators.
 
 Quarkonium annihilation and production happen at the scale $m$: at this  
 scale $m$, the suitable EFT is NonRelativistic QCD (NRQCD).
 In NRQCD soft  and ultrasoft scales remain  dynamical and  their mixing  may complicate 
 calculations,  power counting and do not allow to obtain a Schr\"odinger
formulation in terms of potentials.   
One can go down one step further and integrate out  the soft scale 
in a matching procedure to the lowest energy EFT,
where only ultrasoft degrees of freedom are dynamical. Such EFT is called potential 
  NonRelativistic QCD (pNRQCD)   \cite{Pineda:1997bj,Brambilla:1999xf,Brambilla:2004jw}.
In this case the matching coefficients encode 
the information on the soft scale and represent the potentials.
pNRQCD is making explicit at the Lagrangian level the expansion in $m v^2/mv$.
It is close to a Schr\"odinger-like description of the bound
state, the bulk of the interaction
being carried by potential-like terms, but non-potential interactions,
associated with the propagation of low-energy degrees of freedom 
($Q\bar{Q}$ colour singlets, $Q\bar{Q}$  colour  octets and low energy gluons),
may  still be   present in general. They 
start  to contribute at  NLO (next-to-leading order)  in the multipole expansion of the gluon fields and are 
typically  related to nonperturbative effects
\cite{Brambilla:1999xf} like gluon condensates.

Quarkonium formation happens at the scale $mv$. 
At the scales $mv$ and $mv^2$  the suitable EFT is  pNRQCD.

In this EFT frame,  it is important to establish when $\lQ$ sets in, i.e. when we have to 
resort to non-perturbative methods.
For low-lying resonances, it is reasonable  to assume 
$m v^2 \simg \lQ$. Then, the  system is weakly coupled and we may rely on perturbation theory,
for instance, to calculate the potential. In this case, we deal 
with weak coupling pNRQCD.
The theoretical challenge here is 
performing higher-order perturbative calculations, resum large logarithms in the ratio of the scales  
and the goal is precision physics.

Given that for system with a small radius precision calculations are possible,
in this case quarkonium may become a
benchmark for our understanding of QCD, in particular the transition 
region between perturbative and nonperturbative QCD, and 
for the precise determination
of relevant Standard Model parameters e.g. the heavy quark masses
$m_c, m_b, m_t$ 
and  $\alpha_{\rm s}$. For example,
using the new CLEO data on radiative $\Upsilon(1S)$ decay, the improved lattice determination of the NRQCD matrix elements
and their perturbative pNRQCD calculation, 
it has been possible to obtain in \cite{Brambilla:2007cz} a determination of $\als$ at the $\Upsilon$ mass 
 $\als(M_\Upsilon(1S))=0.184^{+0.015}_{-0.014}$  giving a value  $\als(M_z)=0.119^{0.006}_{-0.005}$ in agreement 
 with the world average.

In weak coupling pNRQCD  the soft scale is perturbative  and  the potentials
are purely perturbative objects. Nonperturbative effects enter 
energy levels and decay calculations in the form of local or nonlocal 
electric and magnetic condensates 
\cite{Brambilla:1999xj}.  We  still lack a precise 
and systematic knowledge  of such nonperturbative purely glue 
dependent objects and it would be important to have for them 
lattice determinations or data extraction (see e.g. \cite{Brambilla:2001xy})
or calculation in models of low energy QCD. 
Notice that the leading electric and magnetic nonlocal correlators 
(that are gauge invariant quantities) may be related 
to the gluelump masses \cite{Brambilla:1999xf}
and to some existing lattice (quenched) determinations 
\cite{Brambilla:2004jw}.
However, since the nonperturbative contributions  are suppressed in the power 
counting it is possible to obtain good determinations of the masses of the
lowest quarkonium resonances  with purely perturbative calculations
in the cases in which the perturbative series is  convergent 
after that the appropriate subtractions of renormalons have been
performed and  large logarithms in the scales ratios  are resummed  
\cite{Pineda:2001ra}.
The potentials are
 matching coefficients that  undergo renormalization, 
develop a scale dependence and satisfy renormalization
group equations.

The static singlet $Q \bar{Q}$ potential is pretty well known.
The three-loop correction to the static potential is now completely
known: the fermionic contributions to the three-loop 
coefficient~\cite{Smirnov:2008pn} first became available, and more
recently the remaining purely gluonic term has been 
obtained~\cite{Anzai:2009tm,Smirnov:2009fh}. 

The first log related to ultrasoft effects arises at three 
loops   \cite{Brambilla:1999qa} . Such logarithm  contribution at N$^3$LO 
and the single logarithm contribution at N$^4$LO may be extracted respectively 
from a one-loop and two-loop  calculation in the EFT and have been calculated 
in \cite{Brambilla:2010pp,Brambilla:2009bi}.

The perturbative series of the static potential suffers from a renormalon ambiguity 
(i.e. large $\beta_0$  contributions) and from large logarithmic contributions.
The singlet 
static energy,  given by the sum of a constant, the static potential and the ultrasoft 
corrections,
is free from ambiguities of the perturbative series. By resumming the large logs using 
the renormalization  group equations and  comparing it
(at the NNLL) with lattice 
calculations of the static energy one sees 
that the QCD perturbative series converges very nicely 
to and agrees with 
the lattice result in the short range    (up to 0.25~fm) and that no nonperturbative
linear (``stringy'') contribution to the static potential exist \cite{Pineda:2002se,Brambilla:2010pp}.
In particular, the 
recently obtained theoretical expression~\cite{Brambilla:2010pp} 
for the complete QCD static
energy at  NNNLL precision has
been used 
to determine $r_0 \Lambda_{\overline MS}$ by comparison with available lattice
data, where $r_0$ is the lattice scale and $\Lambda_{\overline MS}$
is the QCD scale, obtaining 
$r_0\Lambda_{\overline MS} =0.622^{+0.019}_{-0.015}$  for the zero-flavor case. 
This extraction was previously performed
at the NNLO level (including an estimate at NNNLO) in \cite{Sumino:2005cq}.
The same procedure can be used to obtain a precise evaluation of the
unquenched $r_0 \Lambda_{\overline MS}$ value after short distance unquenched
lattice data for the $Q \overline{Q}$ will appear \cite{Donnellan:2010mx}.

The static octet potential is known up to two loops~\cite{Kniehl:2004rk}, see also 
\cite{Pineda:2011db}.
Relativistic corrections to the static singlet potential
have been calculated over the years and are 
summarized in \cite{Brambilla:2004jw}. 

In the case of $QQq$ baryons, the static potential has been determined up to 
NNLO in perturbation theory    \cite{Brambilla:2009cd}   and recently also on the lattice \cite{Yamamoto:2007pf}.
Terms suppressed by powers of  $1/m$  and $r$ in the Lagrangian have been matched 
(mostly) at leading order and used to determine, for instance, the expected 
hyperfine splitting of the ground state of these systems.

In the case of $QQQ$ baryons, the static potential has been determined up to 
NNLO in perturbation theory \cite{Brambilla:2009cd}
and also on the lattice \cite{Takahashi:2000te}. The transition region from 
a Coulomb to a linearly raising potential is characterized in this case also 
by the emergence of a three-body potential apparently parameterized by only one length. 
It has been   shown that in 
perturbation theory a smooth genuine three-body potential shows up at two loops.

For higher resonances $m v \sim \lQ$. In this case, we deal with 
strongly coupled pNRQCD. We need nonperturbative methods to calculate the potentials
and one of the goal is the investigation of the QCD low energy dynamics \cite{Brambilla:1999ja}.
Then the  potential matching coefficients
are obtained in the form of expectation values of gauge-invariant 
Wilson-loop operators. 
In this case, heavy-light meson pairs and heavy hybrids 
develop a mass gap of order $\lQ$ with respect to the energy of the
$Q\overline{Q}$ pair, the second circumstance 
being apparent from lattice simulations.
Thus, away from threshold, 
the quarkonium singlet field  is the only low-energy dynamical 
degree of freedom in the pNRQCD Lagrangian 
(neglecting ultrasoft corrections coming 
from pions and other Goldstone bosons).
The singlet potential  can be expanded
in powers of the inverse of the quark mass;
static, $1/m$ and $1/m^2$ terms were calculated long 
ago~\cite{Brambilla:2000gk,Brambilla:2002nu}.
They involve NRQCD matching coefficients (containing 
the contribution from the hard scale) and low-energy 
nonperturbative parts given in terms
of static Wilson loops and field-strength insertions in the static
Wilson loop
(containing the contribution from the soft scale).
Such expressions correct and generalize previous finding in the Wilson loop approach
\cite{Eichten:1980mw} 
that were typically missing the high energy parts of the potentials, 
encoded into the NRQCD matching coefficients and containing the 
dependence on the logarithms of $m_Q$, and some of the low energy contributions.
The nonperturbative $QQQ$ potentials (static and relativistic corrections) 
have been  obtained in terms of Wilson loops and field strengths insertions in 
\cite{Brambilla:1993zw} and in the second reference of \cite{Brambilla:2009cd}.

In this regime of pNRQCD, we recover the quark potential singlet model. 
However, here the potentials are calculated in QCD by nonperturbative 
matching. Their evaluation requires calculations on the lattice 
or in QCD vacuum models. For calculations inside different QCD vacuum/string models see \cite{Brambilla:1999ja,Brambilla:1996aq}.
Recent progress includes new, precise lattice calculations 
of these potentials 
obtained using the L\"uscher multi-level algorithm \cite{Koma:2007jq}.

As mentioned, which quarkonium state belongs to which regime is an open issue 
and no clear cut method exist to decide this a priori, in the lack of a direct 
way to determine the quarkonium radius \cite{GarciaiTormo:2005bs}. 
Typically the lowest states $\Upsilon(1S), \eta_b$, $B_c$ and possibly $J\psi$ and $\eta_c$
are assumed to be in the weakly coupled regime (for what concerns the soft scale).

\section{Quarkonium spectra, decays, exotics and production}

In  \cite{Brambilla:2010cs, Brambilla:2004jw,Brambilla:2004wf}.
an enormous   set of the most updated phenomenological applications of the EFT framework  outlined above 
to quarkonium spectra, decays and production is  presented and discussed in relation 
to the experimental data.

Here I can only briefly  recall  some selected results.

The energy levels have been calculated at order $m \als^5$  \cite{Brambilla:1999xj,Kniehl:2002br}.
Decays amplitude   \cite{Kiyo:2010jm,Brambilla:2010cs,Brambilla:2004jw}        
and production and annihilation 
\cite{Beneke:2007pj} have been calculated in perturbation theory at high order.
Since for systems with a small radius the nonperturbative contributions  are suppressed in the power 
counting it is possible to obtain good determinations of the masses of the
lowest quarkonium resonances  with purely perturbative calculations
in the cases in which the perturbative series is  convergent 
(after that the appropriate subtractions of renormalons have been
performed) and  large logarithms in the scales ratios  are resummed.
For example in \cite{Brambilla:2000db}
a prediction of the $B_c$ mass has been obtained. The NNLO calculation 
with finite charm mass effects \cite{Brambilla:2001qk}
predicts  a mass that well matches the Fermilab
measurement  \cite{Brambilla:2010cs}
and the lattice determination \cite{Allison:2004be}.
The same procedure has been applied  at NNLO even for higher states
\cite{Brambilla:2001qk}.
A NLO calculation reproduces in part the $1P$ fine splitting 
\cite{Brambilla:2004wu}.
Including logs resummation at NLL, it is possible to obtain a 
prediction for for the $B_c$ hyperfine separation $\Delta=50 \pm 17 
({\rm th}) ^{+15}_{-12} (\delta \als)$ MeV \cite{Penin:2004xi}
and for  the hyperfine separation between the $\Upsilon(1S)$
and the $\eta_b$ the value of  $41 \pm 11 ({\rm th}) ^{+9}_{-8} (\delta
\als)$ MeV  (where the second error comes from the uncertainty in $\als$) 
\cite{Kniehl:2003ap}. This last value turned out to undershoot the 
experimental measurement of BABAR by about two standard deviations.
This is an open puzzles in theory. NRQCD lattice calculations \cite{Gray:2005ur}
obtains a value close to the experimental one but do not include the one loop 
matching coefficients of the spin-spin term that is large and may give a negative 
correction of up to $-20$ MeV \cite{Penin:2010zz}. Recent lattice calculations
\cite{Hammant:2011bt} aims at including the NRQCD matching coefficients in the NRQCD 
lattice calculation and will help to settle this issue.
Another explanation would be related to the presence of a CP light odd Higgs which mixes
with the  $\eta_b$ \cite{SanchisLozano:2010zza}.

An effective field theory of magnetic dipole transition has been given in 
\cite{Brambilla:2005zw},
allowed magnetic dipole transitions between $c\bar{c}$ and $b \bar{b}$  ground states
have been considered in pNRQCD at NNLO  in \cite{Brambilla:2005zw}.
The results are: $\Gamma(J/\psi \to \gamma \, \eta_c) \! = (1.5 \pm 1.0)~\hbox{keV}$
and $\Gamma(\Upsilon(1S) \to \gamma\,\eta_b)$ $=$  $(k_\gamma/71$ $\hbox{MeV})^3$
$\,(15.1 \pm 1.5)$ $\hbox{eV}$, where  the errors account for uncertainties 
coming from higher-order corrections. The width $\Gamma(J/\psi \to \gamma\,\eta_c)$ 
is consistent with the PDG value. 
The quarkonium magnetic moment is explicitly calculated and turns out to be very small in agreement 
with a recent lattice calculation \cite{Dudek:2006ej};
  the M1 transition of the lowest quarkonium states at relative order $v^2$ turn out 
to be completely accessible in perturbation theory \cite{Brambilla:2005zw}.
A description of the $\eta_c$ line shape has been given in  \cite{Brambilla:2010ey}.
and effective field theory calculation  of electric dipole transitions is currently 
in elaboration \cite{Piotr}.
Using pNRCD and Soft Collinear EFT (SCET) a good description of the $\Upsilon(1S)$
  radiative decay have been obtained \cite{GarciaiTormo:2007qs}.

For what concerns decays,  recently, substantial progress has been made in the evaluation of the 
NRQCD factorization formula at order $v^7$ \cite{Brambilla:2006ph}, in the lattice evaluation of the NRQCD matrix elements \cite{Bodwin:2005gg}, in the higher order perturbative calculation of some NRQCD
matching coefficients \cite{Jia:2011ah,Guo:2011tz}
 and in the new,  accurate data on many hadronic 
and electromagnetic decays \cite{Brambilla:2010cs}. 
The data are clearly sensitive to NLO corrections in the Wilson coefficients 
and presumably also to relativistic corrections. 
Improved  theory predictability would entail  the lattice 
calculation or data extraction of the NRQCD matrix elements and 
perturbative resummation of large contribution in the NRQCD matching coefficients.
 Inclusive decay amplitudes have been calculated in pNRQCD in \cite{Brambilla:2002nu}
 and the number of nonperturbative correlators appears to be sizeably reduced with respect to NRQCD
so that new, model independent predictions have been made possible \cite{Brambilla:2001xy}.
Still, the new data on hadronic transitions and hadronic decays pose interesting 
challenging to the theory.
Exclusive  decays mode are more difficult to be addressed in theory 
\cite{Soto:2011id,He:2010pb,Vairo:2003gh}.

For the excited states masses away from threshold, phenomenological applications of the QCD potentials 
obtained in \cite{Brambilla:2000gk} are ongoing \cite{Laschka:2011zr}.
For a full phenomenological description of the spectra and decays it would be helpful to have updated,
more precise and unquenched  lattice calculation of the Wilson loop field strength 
insertions expectation values and of the local and nonlocal gluon correlators \cite{Brambilla:2004jw}.
For recent lattice results on the spectroscopy see \cite{Gregory:2010gm}.

In the most interesting region, the region close to threshold where many
new states, conceivably of an exotic nature have been recently discovered, 
a full  EFT description has yet been constructed nor the appropriate degrees of 
freedom clearly identified   \cite{Brambilla:2008zz,Brambilla:2010cs}.
 An exception is constituted by the $X(3872)$ that 
displays universal characteristics related to its being  so close to threshold, 
reason for which a beautiful EFT  description  could be   obtained 
\cite{Braaten:2003he,Braaten:2009zz}. 

The threshold region remains troublesome
also for the lattice,  althought  several excited states calculations have been recently being 
pionereed.

Lattice results about the crosstalk of the static potential with a pair of 
heavy-light mesons in the lattice have recently appeared   \cite{Bali:2009er}
 but further investigations appear to be necessary.
Several model approaches and predictions for the exotics properties are summarized and described 
in \cite{Brambilla:2010cs}. For a sum rules review see \cite{Nielsen:2009uh}.
The recent discovery at BELLE of two new  exotic charged bottomonium-like resonances 
\cite{Collaboration:2011gj} suggests that many new exotics states will be soon discovered
\cite{Voloshin:2011qa}.

The field of quarkonium production has seen terrific progress in the last 
few years both in theory and in experiments, for a review see  
\cite{Brambilla:2010cs,Brambilla:2004wf,Bodwin:2010py,Lansberg:2008zm,delValle:2011fw}.
Particularly promising seem to be the recent full NLO NRQCD calculation of $J/\psi$ photoproduction
\cite{Butenschoen:2009zy} 
and hadroproduction \cite{Butenschoen:2010rq,Ma:2010yw}, the consequent phenomenological 
applications to the study of $J/\psi$ production at Hera, Tevatron, RHIC and LHC \cite{Butenschoen:2010px}
with the possibility to extract the color octet matrix elements from the combined fits.
The quarkonium polarization remains a very hot topic with theory predictions and approaches
 \cite{Faccioli:2011zz,Brambilla:2010cs}
to be soon validated at the LHC. A calculation of triply heavy baryons production at LHC 
just appeared \cite{Chen:2011mb}.

\section{Quarkonium in media}
The suppression of quarkonium production in
the hot medium remains one of  the cleanest and most relevant
probe of deconfined matter.

However, the use of quarkonium yields as a hot-medium
diagnostic tool has turned out to be quite challenging
for several reasons.
Quarkonium production has already been found to
be suppressed in proton-nucleus collisions by
cold-nuclear-matter effects, which themselves require
dedicated experimental and theoretical attention. Recombination effects
may  play an additional role and  thus transport properties may become 
relevant to  be considered. Finally,  the heavy quark-antiquark interaction
at finite temperature $T$ has to be obtained from QCD.

For observables only sensitive to gluons and light quarks,
a very successfull EFT called Hard Thermal Loop (HTL)
effective theory has been derived  in the past
by integrating out the hardest momenta  proportional to $T$
from the dynamics.  However,   considering also heavy quarkonium in the hot 
QCD medium,  one has to consider in addition to the thermodynamical scales 
in $T$ also the scales of the nonrelativistic bound state and the situation 
becomes more complicate. 

In the last few years years, there has been a remarkable 
progress in constructing EFTs  for quarkonium at finite temperature and 
in rigorously defining 
the quarkonium potential.
In \cite{Laine:2006ns,Laine:2007qy}, the static potential was calculated 
in the regime $T \gg 1/r \simg m_D$, where $m_D$ is the Debye mass and 
$r$ the quark-antiquark 
distance, by performing an analytical continuation of the 
Euclidean Wilson loop
to real time. The calculation was done in 
the weak-coupling resummed perturbation
theory. The imaginary part of the gluon self energy gives an imaginary part to
the static potential and hence a thermal width to
the quark-antiquark bound state. In the same framework, the dilepton
production rate for charmonium and bottomonium was calculated in 
\cite{Laine:2007gj,Burnier:2007qm}.
In \cite{Beraudo:2007ky}, static particles in real-time formalism were 
considered  and the potential for distances $1/r \sim m_D$ was derived 
for a hot QED plasma. 
The real part of the static potential was found to agree with the
singlet free energy and the damping factor with the one found in 
\cite{Laine:2006ns}.
In \cite{Escobedo:2008sy}, a study of bound states in a hot QED 
plasma was performed in a non-relativistic EFT framework. 
In particular, the hydrogen atom was studied for temperatures ranging from 
$T\ll m\alpha^2$ to $T\sim m$, where the imaginary part of the
potential becomes larger than the real part and the hydrogen ceases to exist. 
The same study has been extended to muonic hydrogen in 
 \cite{Escobedo:2010tu}, providing a method to estimate  the effects of a 
finite charm quark mass on the dissociation temperature of bottomonium.

An EFT framework in real time and weak coupling for quarkonium at finite
temperature was developed in \cite{Brambilla:2008cx}   working in real time and 
in the regime 
of small coupling $g$, so that $g T \ll T$  and  $v \sim \als$, which is expected to be valid 
for tightly bound states: $\Upsilon(1S)$, $J/\psi$, ...~.

 Quarkonium in a medium is characterized by different energy and momentum scales;  there are the scales of the non-relativistic bound state  that we have  discussed at the beginning, and there are the  
thermodynamical scales: the temperature $T$, the inverse of the screening 
length of the chromoelectric interactions, i.e. the Debye mass $m_D$ and lower scales, which 
we will neglect in the following.

If these  scales are hierarchically ordered,  then we may expand physical observables in the 
ratio of such scales. If we separate explicitly the contributions from the different scales
at the Lagrangian level this amounts to substituting QCD with a hierarchy of EFTs, which are equivalent 
to QCD order by order in the expansion parameters.
As it has been described in the previous sections
 at zero temperature the EFTs
that follow from QCD by integrating out the scales $m$ and $mv$ are called respectively 
Non-relativistic QCD (NRQCD) and potential NRQCD (pNRQCD).
We assume that the temperature is high enough that $T \gg gT \sim m_D$ holds 
but also that it is low enough for $T \ll m$ and $1/r \sim mv \simg m_D$ to be satisfied, 
because for higher temperature  the bound state ceases to exist.
Under these conditions some possibilities are in order. If $T$ is the next relevant scale after 
$m$, then integrating out $T$ from NRQCD leads to an EFT that we may name NRQCD$_{\rm HTL}$, because 
it contains the hard thermal loop (HTL) Lagrangian
 \cite{Braaten:1989mz}.
Subsequently integrating out 
the scale $mv$ from NRQCD$_{\rm HTL}$ leads to a thermal version of pNRQCD that we may call 
pNRQCD$_{\rm HTL}$. If the next relevant scale after $m$ is $mv$, 
then integrating out $mv$ from NRQCD leads to pNRQCD. If the temperature is larger than $mv^2$, 
then the temperature may be integrated out from pNRQCD leading to a new version of pNRQCD$_{\rm HTL}$  \cite{Vairo:2009ih}.
 Note that, as long as the temperature is smaller than the scale being 
integrated out, the matching leading to the EFT may be performed putting the
 temperature to zero.

The derived potential $V$  describes the real-time evolution of a quarkonium state
in a thermal medium. At leading order, the evolution is governed 
by a Schr\"o\-din\-ger equation. In an EFT framework,
the potential follows naturally from integrating out all  contributions coming from modes
with energy and momentum larger than the binding energy.
For $T < V$ the potential is simply the Coulomb potential. Thermal corrections 
affect the energy and induce a thermal width to the quarkonium state; these 
may be relevant to describe the in medium modifications of quarkonium at low temperatures.
For $T >V$ the potential gets thermal contributions, which are both real and imaginary.

General findings in this picture are:
\begin{itemize}
\item{}  The thermal part of the potential has a real and  an imaginary part. 
The imaginary part of the potential smears out the bound state 
peaks of the quarkonium spectral function, leading to their dissolution prior
to the onset of Debye screening in 
the real part of the potential (see, e.g. the 
discussion in \cite{Laine:2008cf} and applications in \cite{Margotta:2011ta,Miao:2010tk}).
So quarkonium dissociation appears to  be a consequence of the 
appearance of a thermal decay width rather than being due to the color screening of
the real part of the potential; this follows from the observation that the
thermal decay width becomes as large as the binding energy at a temperature 
at which color screening may not yet have set in.
\item{} Two mechanisms contribute to the thermal decay width: the imaginary part of the gluon self energy  induced by the Landau-damping phenomenon (existing also in QED)  
\cite{Laine:2006ns}  and the quark-antiquark color singlet to color 
octet thermal break up (a new effect, specific of QCD)  \cite{Brambilla:2010xn}.
 Parametrically, the first mechanism dominates for temperatures 
such that the Debye mass $m_D$ is larger than the binding energy, while the latter 
dominates for temperatures such that $m_D$  is smaller than the binding energy.
\item{} The obtained singlet thermal potential, $V$, is neither the color-singlet quark-antiquark free energy  nor the internal energy. It has an imaginary part and may contain divergences
that eventually cancel in physical observables \cite{Brambilla:2010xn}.
\item{} Temperature effects can be other than screening, typically they  may appear as
power law corrections or a logarithmic dependence \cite{Brambilla:2010xn,Escobedo:2008sy}.
\item{} The dissociation temperature goes parametrically as  $\pi T_{\rm melting} \sim m g^{4\over 3}$ \cite{Escobedo:2008sy,Laine:2008cf}.
\end{itemize}

The EFT  provides a clear definition of the potential and a  coherent and systematical setup
to calculate masses and widths of quarkonium at finite temperature.
 In \cite{Brambilla:2010vq} 
heavy quarkonium energy levels and decay widths in a quark-gluon plasma, 
below the melting temperature at a
temperature T and screening mass $m_D$ satisfying the hierarchy  
$m \als  \gg \pi T  \gg m \als^2 \gg m_D$, have been calculated  at order $m \als^5$.
This  situation is relevant for bottomonium $1S$   states ($\Upsilon(1S)$, $\eta_b$) at the LHC. 
It has been found \cite{Brambilla:2010vq} 
 that: at leading order the quarkonium masses increase quadratically with $T$ 
which in turn implies the same functional increase in the energy of the dileptons produced in the electromagnetic decays;  a thermal correction proportial to $T^2$ appears in the electromagnetic quarkonium decay  rates;  at leading order a decay width linear with the temperature is developed which implies a tendency to dissolve by decaying to the continuum of the colour-octet states. 

In \cite{Brambilla:2011mk} the leading-order thermal corrections to the spin-orbit potentials of pNRQCD$_{\rm HTL}$
has been calculated and it has been  shown how Poincar\'e invariance is broken by the presence of the  medium.  In \cite{Escobedo:2011ie}
the propagation of a nonrelativistic bound state moving  across a homogenoeus thermal bath have been studied.

In \cite{Brambilla:2010xn,Burnier:2009bk} the Polyakov loop  and the correlator of two Polyakov loops 
at finite temperature  has ben calculated at next-to-next-to-leading order in the weak coupling 
regime and at quark-antiquark distances shorter than the inverse of the temperature and for 
Debye mass larger than the Coulomb potential.  
The calculation has been performed also the in EFT framework 
\cite{Brambilla:2010xn} and a relation between the Polyakov loop correlator and the  singlet and octet  quark-antiquark correlator has been established in this setup.

First attempts to generalize this new picture to the nonperturbative 
regime have been undertaken in \cite{Rothkopf:2009pk}.

\section{Outlook}
Heavy quarkonium physics is a very broad field  spanning many experiments at different facilities around the world and
a very broad set of topics and approaches in theory. The field is in rapid evolution and the experiments at LHC will boost its 
progress. For a summary of the most crucial developments and suggested directions for further advancement I refer you to 
the 65 priorities list at the end of the QWG  document \cite{Brambilla:2010cs}.

\end{document}